\documentclass[aps,prb,reprint,showpacs,superscriptaddress,floatfix]{revtex4-1}
\usepackage{comment}
\usepackage{graphics,amssymb,amsmath,epsfig,color,textgreek}
\usepackage{graphicx}
\usepackage[dvipsnames]{xcolor}
\usepackage{dcolumn}
\usepackage{bm}
\usepackage[colorlinks=true,citecolor=cyan]{hyperref}
\hypersetup{colorlinks=true,citecolor=cyan,linkcolor=red,urlcolor=magenta}
\usepackage{braket}
\usepackage[normalem]{ulem}
\usepackage{cancel}

\usepackage{diagbox}
\usepackage{lipsum}
\usepackage{float}
\usepackage{cleveref}
\usepackage[colorinlistoftodos]{todonotes}
\usepackage{soul}

\crefname{figure}{Fig.}{Figs.}
\crefformat{equation}{Eq.~#2#1#3}

\usepackage[dvipsnames]{xcolor}

\newcommand{\op}[1]{\mathbf{#1}}

\newcommand{\bi}{\begin{itemize}}
	\newcommand{\ei}{\end{itemize}}
\newcommand{\be}{\begin{equation}}
	\newcommand{\ee}{\end{equation}}
\newcommand{\ba}{\begin{eqnarray}}
	\newcommand{\ea}{\end{eqnarray}}

\begin{document}
	
	\title{Probing Bound State Relaxation Dynamics in Systems Out-of-Equilibrium on Quantum Computers}
	
	\author{Heba A. Labib}
	\email{halabib@ncsu.edu}
	\affiliation{Department of Physics, North Carolina State University, Raleigh, North Carolina 27695, USA}
	
	\author{Goksu Can Toga}
	\email{gctoga@ncsu.edu}
	\affiliation{Department of Physics, North Carolina State University, Raleigh, North Carolina 27695, USA}
	
	\author{J.~K.~Freericks}
	\email{james.freericks@georgetown.edu}
	\affiliation{Department of Physics, Georgetown University, Washington DC 20057, USA}
	
	\author{A.~F.~Kemper}
	\email{akemper@ncsu.edu}
	\affiliation{Department of Physics, North Carolina State University, Raleigh, North Carolina 27695, USA}

	\date{\today{}}
	
	\begin{abstract}
		Pump-probe spectroscopy is a powerful tool for probing response dynamics of quantum many-body systems in and out-of-equilibrium. Quantum computers have proved useful in simulating such experiments by exciting the system, evolving, and then measuring observables to first order, all in one setting. Here, we use this approach to investigate the mixed-field Ising model, where the longitudinal field plays the role of a confining potential that prohibits the spread of the excitations, spinons, or domain walls into space. We study the discrete bound states that arise from such a setting and their evolution under different quench dynamics by initially pumping the chain out of equilibrium and then probing various non-equal time correlation functions. Finally, we study false vacuum decay, where initially one expects unhindered propagation of the ground state, or true vacuum, bubbles into the lattice, but instead sees the emergence of Bloch oscillations that are directly the reason for the long-lived oscillations in this finite-size model. Our work sets the stage for simulating systems out-of-equilibrium on classical and quantum computers using pump-probe experiments without needing ancillary qubits. 
	\end{abstract}

	\maketitle

	\section{Introduction}
	
	\begin{figure*}[t]
		\hspace*{-0.4cm} 
		\includegraphics[width=1.0\textwidth]{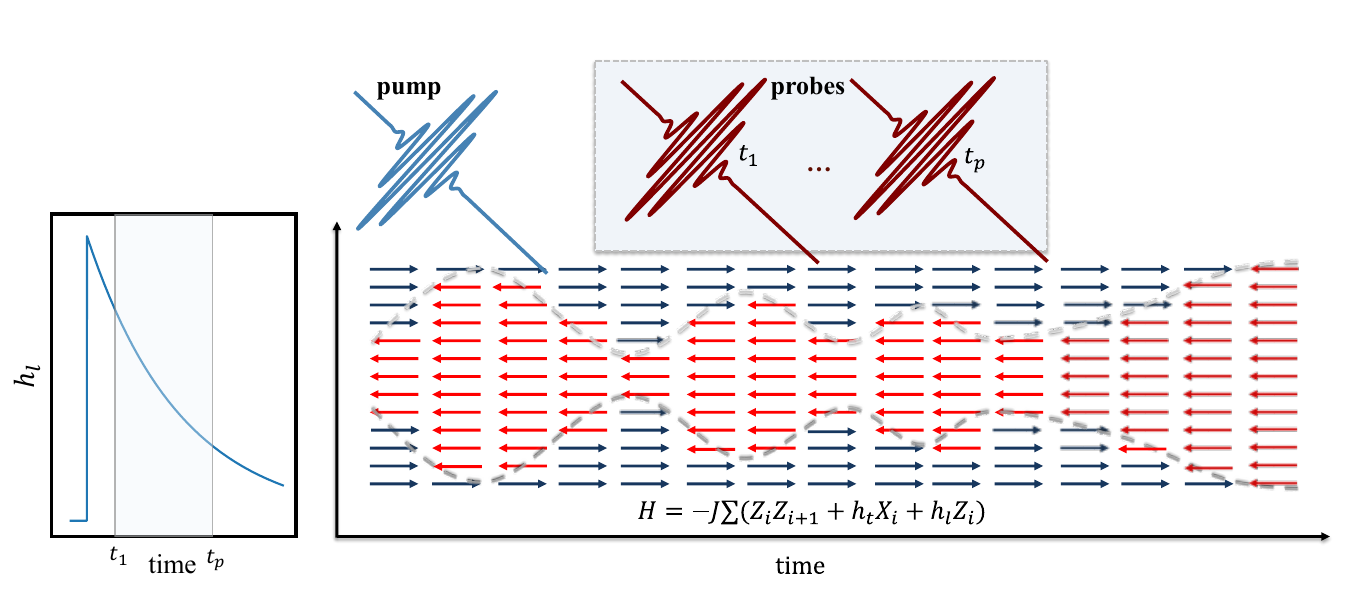}
		\caption{\textbf{Tracking Bound State Evolution by a Pump-Probe Experiment.} 
			We initiate our model in one of the lowest energy states of the final Hamiltonian, either the true vacuum (red) or the false vacuum (blue), and perform both an integrable quench and a non-integrable ramp that relaxes over time. We measure $\langle Z_0(t) \rangle $ and probe the relaxation dynamics where $t_p$ is the time we hit our system with a probe pulse. By computing $\langle [Z_0(t), Z_0(t_p)]\rangle $ using the functional derivatives approach, we observe the change in the number of bound states (mesons), corresponding to the energy of the bound domains/bubbles of false (blue) or true (red) vacua. We also track the emergence of Bloch frequencies due to the presence of the lattice as the parameters of the Hamiltonian relax back to equilibrium.
		}
		\label{fig:mainfig}
	\end{figure*}
	
	The primary method for experimentally interrogating many-body systems, such as those that arise in quantum materials and in quantum simulators, is by measuring dynamical
	observables --- time or frequency-dependent correlation functions --- that reveal the excitations out of the ground state,
	and provide a fingerprint of the many-body system and its ground
	state.  For example, multiple models give rise to a ferromagnetic ground state, but the excitation spectrum is markedly different.
	Dynamical correlation functions are typically treated
	in a linear response formalism; each correlation function $\chi^R(t,t')$
	represents the response in an observable $\mathbf{A}(t)$ to some weak external perturbation $\phi(t') \mathbf{H}'$, which can be treated at linear order through perturbation theory\cite{mahan,bruus,freericks2021two}.

	A different approach to studying many-body systems is to drive them out of equilibrium. Systems out of equilibrium 
	exhibit rich and complex dynamics that challenge conventional theories\cite{maldonado_theory_2017,stefanucci_nonequilibrium_2013, 2009Tsuji,xu_unconventional_2022}. 
	The advancement of experimental techniques in this arena, particularly ultrafast pump-probe spectroscopy, has allowed scientists to probe these out-of-equilibrium phenomena\cite{Bloch:2008,2014Krull,2016Shao, cabanillas2011pump, fushitani2008applications,zonno2021time, perfetto2020time}, revealing new states of matter and emergent behaviors. 
	Non-equilibrium dynamics provide new insight into a wide range of phenomena, including superconductivity\cite{revelle2019theory,2015Orenstein}, magnetism, and quantum transport\cite{mazza_suppression_2019,karamlou2022quantum}.
	This shift in focus has also opened up a wealth of new questions regarding thermalization\cite{riera2012thermalization,robinson_signatures_2019}, dissipation\cite{xu_unconventional_2022}, and the formation of novel phases\cite{2017Kennes,osborne2023probing}, driving further exploration of non-equilibrium phenomena in condensed matter physics\cite{stefanucci_nonequilibrium_2013}. 
	
	In a previous work\cite{kokcu_linear_2024}, we proposed a linear response-based
	quantum simulation approach
	to obtain dynamical correlation functions on quantum computers or simulators that rely on a direct mapping between the experiment and the quantum simulation; in essence, the experimental procedure is simulated and correlation functions are obtained with no  ancillary qubits. Here, we generalize this approach to systems that are out of equilibrium, leveraging the strength of the linear response framework to obtain non-equilibrium dynamics as well. 
	
	We demonstrate this technique by investigating the bound-state dynamics in the mixed-field Ising model after a time-dependent quench. We measure the  longitudinal magnetization to obtain non-equilibrium correlation functions along a time-dependent perturbation, showing that non-equilibrium (pump-probe) dynamics can be captured in this framework and reveal the underlying physics of the system.
	
	The mixed-field Ising model has many interesting properties and connections to many different fields of physics. 
	Notably, it provides a minimal model for the ``false vacuum'' scenario that arises in quantum field theory and cosmology, 
	where the system can get stuck in a meta-stable state that lives up to billions of years but eventually decays\cite{kormos_real-time_2017}.
	
	The transition from a system's local minimum, a false vacuum, to its ground state, or the 'true' vacuum, is made possible through bubble nucleation\cite{lagnese_detecting_2023, luo2025quantum} --- the formation of clusters, or bubbles, of the stable, true vacuum states, in regions in the universe/system where these states can form. The true vacuum eventually engulfs the universe through the bubble growth propelled by the lower energy of the true vacuum. The bubbles
	formed throughout the 1-D chain are bound states of the
	system, to which we refer to as mesons, or mesonic masses, in analogy to
	the bound states formed in quantum chromodynamics
	(QCD) through the quark-antiquark pairs being held together, mesons in the Ising model are kink-antikink pairs
	bound by a linear potential that we will introduce in the coming sections.
	The study of such low-energy spectra of such systems can open many doors to explore non-equilibrium processes, especially dynamical phase transitions in classical and quantum field theories\cite{coldea_quantum_2010,osborne2023probing}.

	In the ferromagnetic regime of the mixed-field Ising model,
	the distinction between the two vacua is provided by the
	longitudinal field, depending on whether the longitudinal field is parallel (true vacuum) or anti-parallel (false vacuum) to the magnetization. The vacua also have distinct dynamics; the true vacuum exhibits \textit{confining} dynamics, whereas the false vacuum exhibits \textit{anti-confining} dynamics.  We will outline these below.
	
	Previous works have studied this system using quench spectroscopy,
	where one or more of the parameters are rapidly changed and the propagation of the system in the new model is observed\cite{delfino_theory_2017, fonseca_ising_2006,gritsev_spectroscopy_2007, krasznai_schrodinger_2024, tan_observation_2021, bhaseen_aspects_2005}. It has also been experimentally demonstrated that this setup is implementable on a trapped-ion quantum computer to study bubble nucleation as the longitudinal field crosses the critical point $h_c = 0$  and track how the metastable state decays into a stable one by effectively simulating a first-order phase transition.~\cite{luo2025quantum}
	
	In addition to field theory, a similar setup also appears in material science. For instance, several spectroscopic experiments have investigated the massive spectra or the $E_8$ spectrum \cite{zou20218, kjall2011bound} in various quantum materials \cite{watanabe_exploring_2024, Amelin2022, sim_nonlinear_2023,wang2015spinon,morris2014hierarchy,grenier2015longitudinal, zou20218}. In particular, studies on a ferromagnetic Ising chain material $\text{CoNb}_2\text{O}_6$ have revealed that by tuning the confining field $h_l$ via inelastic neutron spectroscopy, one can switch between weak confinement and strong confinement regimes, showing that these models can be accurately captured in line with confinement mechanisms.

	In a pump-probe experiment, the application of an ultrafast pump pulse injects energy into the system. After being driven far from equilibrium, this added energy is not retained indefinitely, but instead, it dissipates into an implicit, infinite, background over time. In our study, we adopt this viewpoint by encoding the dissipation dynamics with our change of parameters following a parameter "quench". Under such circumstances, we aim to probe the relaxation dynamics of the non-integrable mixed-field Ising model.
	A schematic of the method is shown in \cref{fig:mainfig}; where the pump-probe experiment, consisting of a pump pulse and a stream of probe pulses to convey the measurement process, is depicted.
	Our measurements, which are just local magnetization measurements,  will show the underlying (anti-)confinement dynamics by revealing the bound states and Bloch oscillations. 
	Through the linear response method, we obtain the necessary non-equal time two-point correlators via local measurements to study these dynamics, observing the typical bound states and Bloch oscillations that occur in this model, but also revealing their temporal dynamics as the system returns to equilibrium.

	\section{Confined Quasi-particles in the Mixed Field Ising Hamiltonian}\label{sec:Ising_section}

	To set the stage for our discussion of non-equilibrium correlation functions, we briefly review the situation in equilibrium in the two relevant scenarios: true and false vacua, with an excitation that belongs to the other vacuum.
	
	The mixed-field Ising Hamiltonian is given by 
	\begin{align}\label{eq:mfim}
		\mathbf{H}_{\rm MFIM} = -J \sum (Z_i Z_{i+1} + h_t X_i + h_l Z_i)
	\end{align}
	
	which serves as a one-dimensional model to study false vacuum decay\cite{lagnese2021false}. We set $J=1$ and $h_t = 0.25$. We simulate the dynamics of this model using exact diagonalization of a $N=12$ site-chain with periodic boundary conditions. While the Ising model with both fields turned off has two degenerate ground states with opposite magnetization, turning on the transverse field results in the well-known phase diagram with a phase transition at $h_t^{\rm critical}=J$; below this critical value, in the ferromagnetic phase, the excitation dynamics are dominated by domain wall propagation in the ferromagnet.~\cite{calabrese_quantum_2012-1, pfeuty1970one}  Further, by turning on a longitudinal field $h_lZ_i$ in the ferromagnetic phase, the degeneracy in the ground state is lifted as the $Z_2$ symmetry is broken. 
	The system now has two distinct low-lying states: a ``true vacuum'' state and a ``false vacuum'' state; we assign the terms referring to the direction of the longitudinal field.

	In what follows, we will consider the dynamics of bubbles of one type of vacuum in the background of the other.  In equilibrium, these can readily be formed by vacuum tunneling~\cite{rutkevich1999decay}. Alternatively, rapid quenches of the Hamiltonian parameters can produce bubbles, which is the situation we will consider here.
	
	\begin{figure}[t]
		\hspace*{-0.3cm} 
		\includegraphics[width=0.51\textwidth]{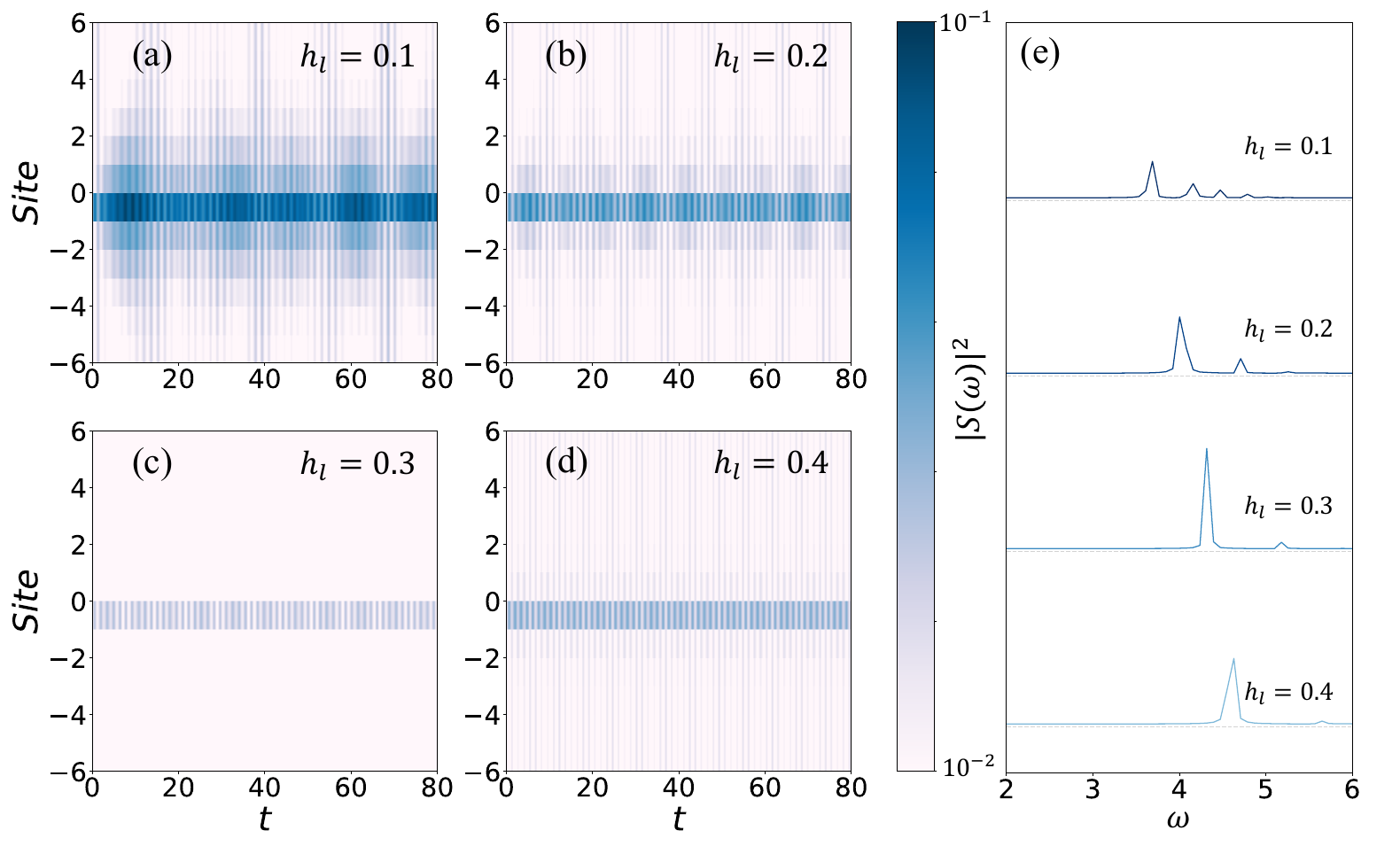}
		\caption{\textbf{Confining Dynamics in Equilibrium.} Confined quasi-particle propagation in the mixed-field Ising model after a transverse field quench $h_t=0.0\rightarrow h_t =0.25$ at $t=0$ together with a longitudinal field quench from $h_l=0.0$ to the values as shown in the panels. \textbf{(a)}-\textbf{(d)} show the confined propagation in $\langle \psi_0|\op{Z_i}(t)\op{Z_j}(t)|\psi_0\rangle_c $. Increasing $h_l$ results in increased confinement. \textbf{(e)} Fourier transform of $\langle \op{Z_i}(t)\op{Z_j}(t)\rangle_c$. The peaks at each $h_l$ value correspond to the bound states or the mesons formed due to the confining field. All panels show $1+\langle \psi_0|\op{Z_i}(t)\op{Z_j}(t)|\psi_0\rangle_c$ on a logarithmic scale.}
		\label{fig:equilibrium}
	\end{figure}
	
	\subsection*{True vacuum with false vacuum bubbles}
	
	If the system starts in the true vacuum state, the domain wall
	excitations get confined as the $h_l Z$ term in the Hamiltonian introduces an attractive potential that inhibits the spreading of the domain walls:
	\begin{align}\label{eq:attractive}
		V(r) = -\chi r
	\end{align}
	where $\chi$ corresponds to the energy gain of flipping one spin to the opposite direction\cite{pomponio_bloch_2022}
	\begin{align}
		\chi = 2|h_l|(1-h_t^2)^\frac{1}{8}.
		\label{eq:chi}
	\end{align}
	The resulting energy cost increases proportionally with respect to the distance between the domain walls. In addition to prohibiting the propagation of the domain wall quasi-particles, the longitudinal field alters the spectrum of the transverse field Ising model. The mesonic masses are the zero-momentum states of the spectrum; and away from zero momentum, the system exhibits an expected dispersion relation with a positive curvature\cite{robinson_signatures_2019}. The above arguments are valid as long as the transverse field is not close to the critical value, i.e., $h_t < h_t^{\rm critical}$; then, the quasi-particle dynamics are effectively described by the two-kink approximation \cite{rutkevich_weak_2010,rutkevich_kink_2018}.

	Under the condition that the quench applied is weak, the system remains close to its ground state; therefore, the post-quench dynamics show the bound state spectrum. 
	This can be probed by measuring order parameters such as the total magnetization $M(t)=\sum_i \langle \op{Z_i}(t) \rangle$, or \emph{equal}-time correlation functions such as $\langle \op{Z_i}(t) \op{Z_j}(t) \rangle$ \cite{delfino_correlation_2018,bhaseen_aspects_2005,lagnese_confinement_2020}. As a demonstration, we show the connected correlation function 
	
	\begin{align}
		\langle \op{Z_i}(t) \op{Z_j}(t) \rangle_c  =\langle \op{Z_i}(t) \op{Z_j}(t) \rangle -\langle \op{Z_i}(t)\rangle \langle \op{Z_j}(t) \rangle
	\end{align}
	for various values of the quenched longitudinal field $h_l$ in \Cref{fig:equilibrium}.
	We see that increasing $h_l$ produces stronger confinement (panels a-d),
	in agreement with \cref{eq:chi}. To get the spectrum, we take the Fourier transform of these correlators, which is shown in \cref{fig:equilibrium} (e). As the longitudinal field increases, the peaks in the spectrum, indicating the bound state formation, move towards higher values of $\omega$, showing that the energy penalty of creating false vacuum domains increases, and consequently, the energies of the bound states increase.
	
	\begin{figure}[b]
		\hspace*{-0.45cm} 
		\includegraphics[width=0.51\textwidth]{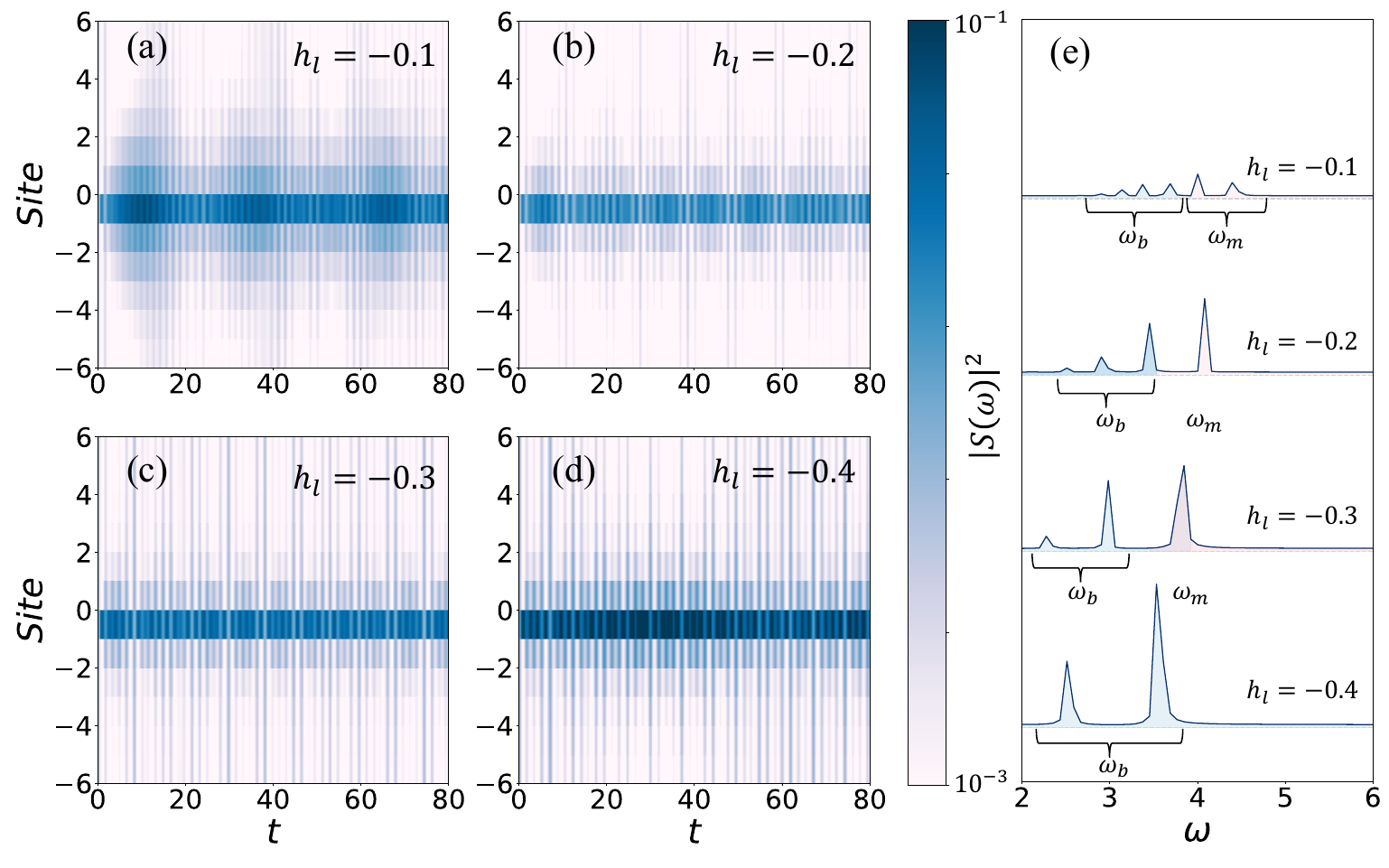}
		\caption{\textbf{Anti-confining Dynamics in Equilibrium.} Quasi-particle propagation in the mixed-field Ising model after a transverse field quench $h_t=0.0\rightarrow h_t =0.25$ at $t=0$ together with a longitudinal field quench from $h_l=0.0$ to values as shown in the panels. \textbf{(a)}-\textbf{(d)} show the propagation in $\langle \psi_0|\op{Z_i}(t)\op{Z_j}(t)|\psi_0\rangle_c$ under an anti-confining field quench $h_l < 0$. Increasing $|h_l|$ gives rise to a number of Bloch frequencies below the threshold of the bound states. We note here that the Bloch oscillations are seen as the envelope of the oscillations across the lattice, while the internal oscillations pertain to the domain walls forming a bound state. \textbf{(e)} Fourier transform of $\langle \op{Z_i}(t)\op{Z_j}(t)\rangle_c$. The peaks at each $h_l$ value correspond to the higher harmonics of the Bloch frequencies $\omega_b$ (in shaded blue) as well as the disappearance of the mesonic (bound) contributions (in shaded pink) to the spectrum. All panels show $1+\langle \psi_0|\op{Z_i}(t)\op{Z_j}(t)|\psi_0\rangle_c$ on a logarithmic scale.
		}
		\label{fig:equilibriumfv}
	\end{figure}
	
	\subsection*{False vacuum with true vacuum bubbles}
	
	In the case of a false vacuum, we start with the metastable state by inverting the orientation of the longitudinal field. Energetically, false vacuum decay into the ground state is favored, and naively, one might expect that a true vacuum bubble to grow until it fills the system; for this reason, this scenario is termed ``anti-confining.''
	However, these excitations are also confined, but here due to the presence of the lattice; the freely-propagating excitations are subject to the discrete lattice confining the unbounded propagation into an oscillatory one.
	
	We repeat the analysis of the previous section for the false vacuum case and calculate the connected correlation function $\langle \op{Z_i}(t)\op{Z_0}(t)\rangle_c$ and its Fourier transform at various $h_l$ values. In \cref{fig:equilibriumfv} (a)-(d), we show the connected correlators; comparing this to the true vacuum case in \cref{fig:equilibrium} (a)-(d), we see that these plots show persistent long lived oscillations especially for higher values of the quench field. The amplitude oscillations (the envelope) can also be seen in the spectrum \cref{fig:equilibriumfv} (e), where they lead to the existence of additional peaks compared to the true vacuum case.

	While we see a band of bound states in the range $\omega = [3.5,4.5]$, extra oscillations are seen in the frequency range just below that of the mesons and are known as Bloch oscillations, and their origin can be traced back to the existence of the lattice, which presents a cutoff at high momenta, leading to a cosine-like periodic band structure and Bloch wavefunctions. Essentially, Bloch oscillations are to occur at the boundaries of the Brillioun zone at $k = \pm \pi$ where the motion reverses.  In \cref{fig:equilibriumfv}, the Bloch oscillations are the translational dynamics along the site axes existing as the amplitude modulation of the internal oscillations pertaining to the bound states. In the frequency domain, the mass of the lowest meson (the smallest bound state) sets a threshold, or the energy gap, for the spectrum; below this threshold, Bloch oscillations dominate the dynamics. Bloch oscillations appear as distinct peaks starting from $\omega_b \approx \chi  $ and a series of equally-spaced higher harmonics  $n \omega_b$ until reaching the mesonic threshold. As the longitudinal field increases, the dynamics of the system shifts from mesonic contributions to the structure of Bloch oscillations in agreement with what we expect for an anti-confining regime.\cite{pomponio_bloch_2022,rutkevich_energy_2008}

	As we will demonstrate below, although the close proximity of these two physically different peaks makes it naturally harder to distinguish, frequency-selective linear response can help to identify Bloch oscillations from mesonic excitations.

	\section{Functional Derivative Approach for Systems out of Equilibrium}

	In this section, we extend our method of obtaining two-point correlation functions through the linear response framework to the study of time-dependent Hamiltonians.

	In a typical pump-probe experiment, the system is first driven out of equilibrium by a strong electromagnetic pulse, followed by a much weaker probe pulse. The pump typically induces a complex and highly non-linear response, while the probe, being weak, perturbs the system only slightly; in \textit{time-resolved angle-resolved photoemission spectroscopy} (TR-ARPES), for example, the pump is responsible for exciting the electrons, and the probe is responsible for the photoemission~\cite{boschini_time-resolved_2024}. As a result, the system’s response to the probe can be accurately described within the framework of linear response theory.

	For simplicity, we model the system's dynamics using an effective Hamiltonian that already incorporates the influence of the pump, and treat the probe as a small time-dependent perturbation~\cite{vernes2005formally,PhysRevA.110.043520}. Accordingly, the dynamics follow the Hamiltonian given as, 
	\begin{align}\label{eq:pumpprobeH}
		\boldsymbol{\mathcal{H}}(t) = \op{H_0}(t)+ \op{H_{probe}}(t)
	\end{align}
	where $\op{H_0}(t)$ describes the effective time-dependent Hamiltonian in response to the pump and $\op{H}_{probe}(t)=\phi(t)\op{H'}$ is the probe Hamiltonian. The system is in equilibrium before the pump arrives at some time (or time interval). Following this, the linear response formalism can be used to find the response of the system due to the probe pulse.
	
	Given that  the perturbation is small, any observable expectation value is dominantly linear in the perturbation:
	\begin{align}
		\delta A(t) &= \langle \psi_0| \op{A}(t) |\psi_0\rangle_{\rm probe} - \langle \psi_0| \op{A}(t) |\psi_0\rangle_{\rm unperturbed} \nonumber \\
		&= \int dt' \chi^R(t,t') \phi(t') + \mathcal{O}(\phi^2).
	\end{align}
	where $\langle \psi_0| \op{A(t)} |\psi_0\rangle_{\rm probe}$ is the expectation value with the influence of the probe and $\langle \psi_0| \op{A(t)} |\psi_0\rangle_{\rm unperturbed}$ is the expectation value in the absence of it. This procedure allows us to isolate the response due to the probe such that $\chi^R(t,t')$ is the expected retarded response function which contains information on the underlying physics of the system. 
	
	By calculating the functional derivative of $\delta A(t)$ with respect to the temporal part of the perturbing field, $\phi(t)$, we can directly obtain the dynamical correlation function $\chi^R(t,t')$:
	\begin{align}\label{eq:functional_derivative}
		\chi^R(t,t') =  -i 
		\theta(t-t')
		\braket{\psi_0 | \left[ \op{A}(t), \op{H'}(t') \right] | \psi_0}.
	\end{align}
	In equilibrium, the $\chi^R(t,t')$ is just a function of the
	time difference $t-t'$, and any Fourier analysis is done with
	respect to this axis.  Out of equilibrium, the pump breaks time
	translation in variance, and this should be accounted for.~\cite{revelle2019theory} One way to do so is to include a
	probe field centered at several times $t_p$, in essence, introducing a new time-axis into the simulation, and taking the Fourier transform along $t$ of $\delta A(t)$
	for the various arrival times of the at different times $t_p$. Doing so, we obtain  
	\begin{align}
		A(\omega, t_p) = \chi^R(\omega, t_p) \phi(\omega) + \mathcal{O}(\phi^2).
	\end{align}
	This allows us to track the dynamics as a function of $t_p$ after the action of the pump, study the relaxation dynamics, and observe how the correlations change\cite{revelle2019theory}. 
	We also highlight that before computing the functional derivative of the observable expectation value, adding a damping factor to $\delta A(t)$ of the form $e^{-t/\tau}$ is advantageous in improving the signal computed and smoothening out the life-times of the quasi-particles.

	We apply this method to the study of the dynamics of the mixed-field Ising model (\ref{eq:mfim}) after a quench, followed by a subsequent time-dependent return to the pre-quench Hamiltonian and choosing $\op{A}(t)=\op{Z}(t)$. This setup is intended to mimic the typical response of a system in a pump-probe experiment,
	where the pump induces some rapid change in the Hamiltonian parameters, and the return to equilibrium as the added energy from the pump dissipates.

	The effect of the pump is to turn on the longitudinal field at $t=0$, from $h_l = 0 $ to $h_l = h_{l_{\rm max}}$, which then relaxes back to zero over time.  This corresponds in the mixed-field Ising model to rapidly quenching into a high-confinement regime, with long-lived oscillations, and then relaxing back to a low-confinement one.
	We set the transverse field to $h_t < h_t^{\rm critical}$, guaranteeing that the excitations are spin flips (domain walls) in the z-direction.

	The probe, which may have some frequency structure, arrives centered at
	$t=t_p$ and is of the form:
	\begin{align}
		\op{H}_{\rm probe}(t) = \phi(t) \op{H'} = \phi(t) \op{Z}_0
	\end{align}
	We choose a time range in which the probe consecutively takes place and measure the responses at each respective $t_p$, and investigate this relaxation dynamics subject to a slow ramp pump over $h_l(t)$ by obtaining $\chi(\omega,t_p)$.

	\begin{figure}[t]
		\includegraphics[clip=true, trim = 4 0 0 0, width=0.49\textwidth]{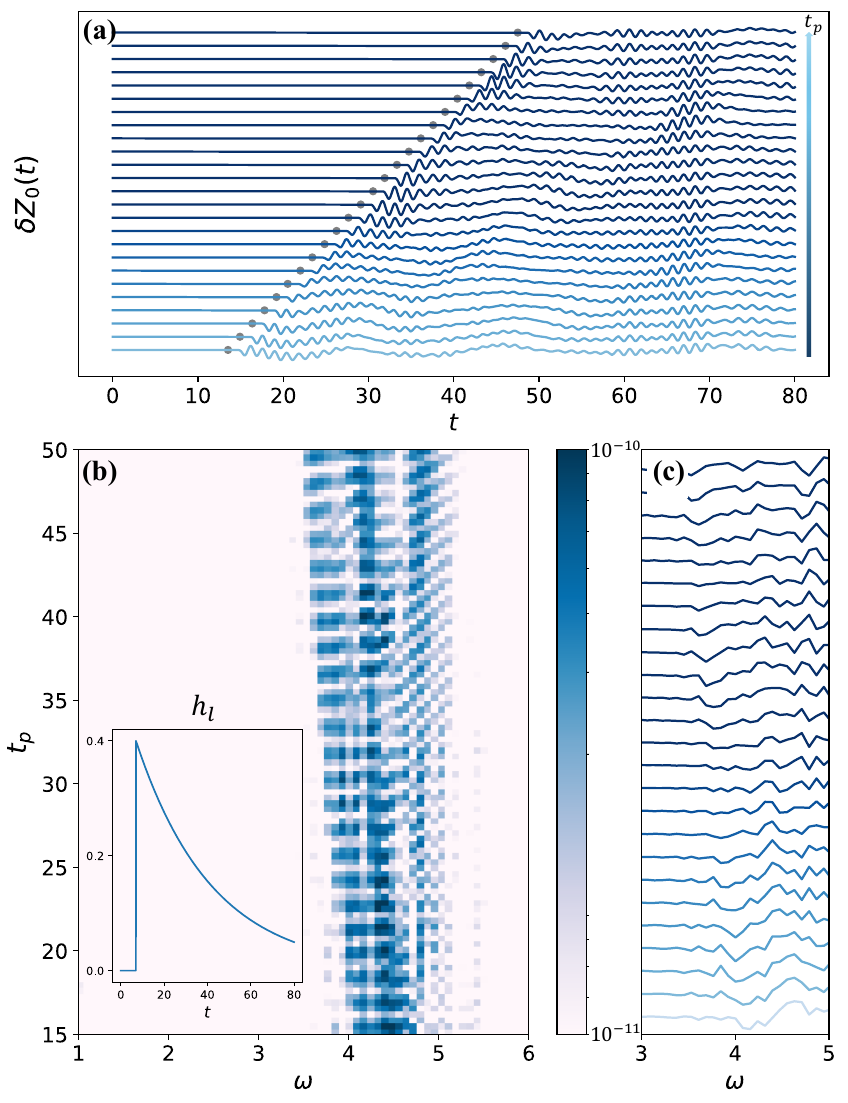}
		\caption{\textbf{True Vacuum Evolution in a confining scenario.} True Vacuum Evolution under a time-dependent Hamiltonian with $h_l$ that goes from $h_l = 0.0 $ to $h_l = 0.4$. \textbf{(a)} Time-domain data for the order parameter $\delta Z_0(t) $ before extracting the correlation functions by the functional derivatives approach. Grey dots indicate the time $t_p$ that the probe field is applied indicating separate experiments and separate measurements. \textbf{(b)} False color plot for $|\chi(t_p,\omega)|^2$. The plot shows the bound-state spectrum as a function of $t_p$ and $\omega$ to study the change in the spectrum as the Hamiltonian changes in time. The inset shows the time-dependence of the longitudinal field $h_l$. \textbf{(c)} Line plot for $|\chi(t_p,\omega)|^2$ (log-scaled) to further analyze the signal peaks trend.}
		\label{fig:tv}
	\end{figure}
	\section{Results}
	\subsection{True Vacuum Analysis}

	\begin{figure}[t]
		\hspace*{-0.5cm}
		\includegraphics[width=0.51\textwidth]{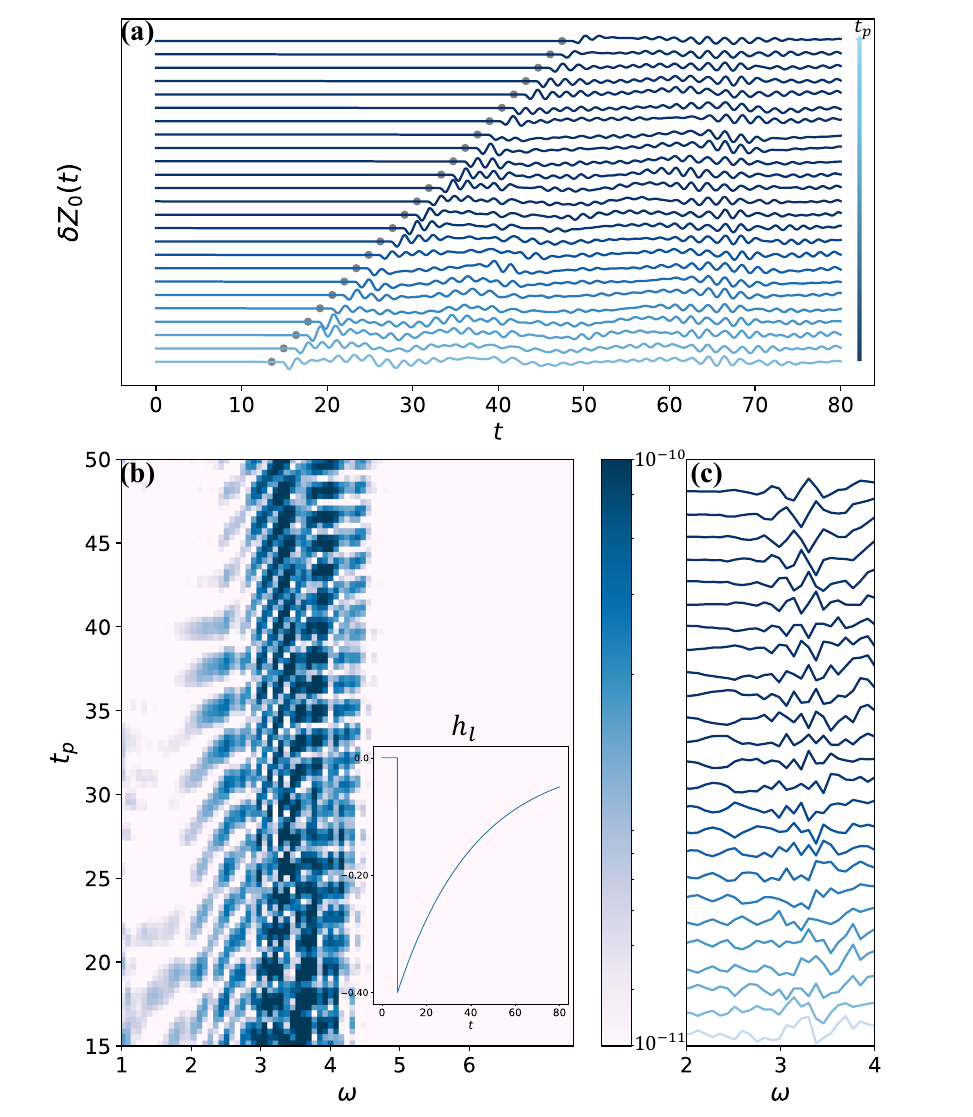}
		\caption{\textbf{False Vacuum Evolution in an anti-confining scenario.} False Vacuum evolution under a time-dependent Hamiltonian with $h_l$ that goes from $h_l=0$ to $h_l = -0.4$. \textbf{(a)} Time-domain data for the order parameter $\delta Z_0(t)$ before extracting the correlation functions by the functional derivatives approach. The probe time $t_p$ is indicated by the gray dots. \textbf{(b)} False color plot for $|\chi(t_p,\omega)|^2$. The plot shows the spectrum as a function of $t_p$ and $\omega$ to study the change in spectrum as the Hamiltonian changes in time. Inset shows the longitudinal field time-dependence. \textbf{(c)} Line plot for $|\chi(t_p,\omega)|^2$ (log-scaled) to zoom in on the signal peaks trend for the corresponding $h_l$ values in \textbf{(b)}. }
		\label{fig:fv}
	\end{figure}

	We start the discussion of our results by focusing on the true vacuum case. 
	Following the recipe laid out in the previous section we choose $\op{H}_0(t)$ such that we start in the true vacuum state; the time profile of the longitudinal field for this case is shown in the inset of \cref{fig:tv} (b) of the form
	\begin{align}\label{eq:fieldform}
		h_l(t) = h_{l_{ \rm max}}\theta(t-t_0) e^{-(t-t_0)/\tau}.
	\end{align}
	where $\theta(t)$ is the Heaviside step function and $t_0$ is when the pump takes place. We also fix the temporal part $\phi(t)$ of the probe field to be a Gaussian with a width $\sigma_t$ small enough to capture all the bound state signatures in the frequency domain; we then probe the system in a time range $t_p \in [15, 50]$ in different separate experiments.

	We time evolve the system and measure $\delta Z_0(t)$, which is shown in \cref{fig:tv} (a) where each line represents a separate simulation with the gray dots indicating the time $t=t_p$ when the probe field is applied. We see that after the probe field arrives $\delta Z_0(t)$ displays persistent, long-lived oscillations similar to the equilibrium case. Before $t_p$, the signal is zero, respecting causality.

	To further investigate these oscillations and to obtain the retarded correlation function we follow the next step in the linear response framework which is to obtain $|\chi(t_p,\omega)|^2$ by calculating the functional derivative of the response $ \delta Z_0(t)$ with respect to $\phi(t)$ in frequency domain;
	the result can be seen in \cref{fig:tv} (b). 
	We also show line cuts on a logarithmic scale in panel (c).

	Probing at $t_p\approx 15.0$ immediately after the pump takes place allows us to observe the dynamics at the high-confinement regime where $h_l \sim h_{l_{\rm max}} = 0.40$. In this case, the bound states are of higher energy, and the dynamics are more confined, as shown by the prominent peaks that are contained in a narrow window between $\omega \approx 4.0 -5.0$ in panels (b) and (c).
	
	At later times, the longitudinal field relaxes towards zero, and the system reaches a low-confinement state, which results in the widening of the narrow excitation window and we start to observe a number of bound states in the low-frequency range.
	
	In \cref{fig:tv} (c), we zoom in on the range where the bound-states lie in the frequency domain. The peaks signify the bound state (or a meson). We see that as $h_l(t)$ relaxes back to its original value, the trend in the bound state frequencies evolves from high frequencies to lower frequencies, and the number of low-energy states increases.

	In addition, the mesonic excitations are also reflected in the dynamics along $t_p$. As such, a Fourier analysis along the $t_p$ axis can reveal spectral fingerprints of both collective excitations, i.e. super-positions of the low-energy excitations (mesons), or the relaxation modes as the system reaches equilibrium over time. We further discuss this in Appendix ~\ref{app:tp}.

	\subsection{False Vacuum Decay and Bloch Oscillations}

	The same analysis can also be carried out for the false vacuum case, in which we start with a pump that produces a longitudinal field that is anti-aligned with the initial state's magnetization of a similar functional form of \cref{eq:fieldform}, but with $h_{l_{ \rm max}} < 0$, thus allowing us to work in the vicinity of false vacuum dynamics. An opposite magnetization to the external field means the system will try to align and reach the true ground state. 
	
	Our results show that, even though reaching the true vacuum state is the energetically favorable process, we still see long-lived bound state dynamics in agreement with the existence of Bloch oscillations in the equilibrium case we discussed in the previous sections.  We will show that these Bloch oscillations can be studied through our pump-probe experiments. 
	
	\begin{figure*}[t]
		\hspace*{-0.6cm} 
		\includegraphics[width=1.05\textwidth]{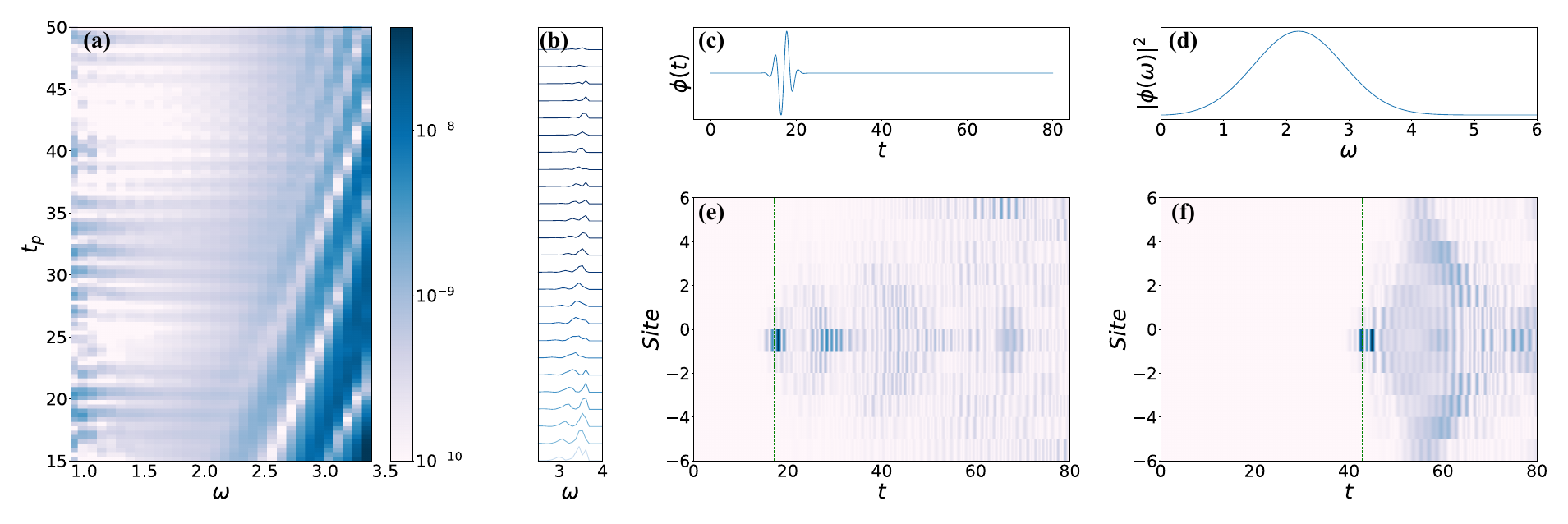}
		\caption{\textbf{Frequency Selectivity to track Bloch Oscillations.} We perform a frequency-selective calculation with a probe of the form $ \phi(t) \propto \sin(\omega(t-t_p))$ to find the higher harmonics Bloch frequencies. \textbf{(a)} False color plot for $|\chi(t_p,\omega)|^2 $under a frequency-selective probe centered at frequency $\omega = 2.2$ and of width $\frac{1}{\sigma}$ where $\sigma = 1.5$. The plot shows the spectrum as a function of $t_p$ and $\omega$ to study the emergence of Bloch frequencies as the Hamiltonian changes in time. We see that the lower $|h_l|$ becomes, the less these frequencies are. For higher $|h_l|$ values, the frequencies become uniformly spaced and one can resolve a number of the higher harmonics that do not overlap with the energy differences. \textbf{(b)} Line plot for $|\chi(t_p,\omega)|^2$ to zoom-in into the signal peaks trend of corresponding $h_l$ values in \textbf{(a)}. \textbf{(c)} Frequency-selective probe in time-domain. \textbf{(d)} Frequency-selective probe in frequency domain. \textbf{(e)} $|\delta Z_i(t)| $ in time-domain under a frequency selective probe centered at $t_p = 17.12$ corresponding to a longitudinal field value of $h_l = -0.296$.
			\textbf{(f)} $|\delta Z_i(t)| $  in time-domain under a frequency selective probe centered at $t_p = 42.92$ corresponding to a longitudinal field value of $h_l = -0.142$.}
		
		\label{fig:bloch}
	\end{figure*}
	In \cref{fig:fv}, we show the full low-energy spectrum for a time-dependent Hamiltonian pumped from $h_l = 0.0$ to $h_l = -0.40$ and relaxing back to zero. As before, by measuring at different $t_p$ points (indicated by the grey dots in \cref{fig:fv} (a)), we can study the trend of the decay and the emergent low-frequency oscillations as a function of time. \cref{fig:fv}(a) shows the magnetization $\delta Z_0(t)$. We see that it exhibits long-lived oscillations with a number of frequencies. Comparing this with the true vacuum case, we can already observe that there are more frequency modes in the false vacuum case, as expected from the existence of Bloch oscillations.

	To obtain the spectrum, we repeat the functional derivative procedure.  In \cref{fig:fv} (c), we show $|\chi(t_p,\omega)|^2$, which shows three regions where we see prominent peaks. The first area near $\omega\approx0.0-1.0$ corresponds to the energy differences of bound states. The second bright area in the region $\omega\approx 2.0-3.0$ is due to the Bloch oscillations, and finally, we have the usual frequencies for the mesonic bound states near $\omega\approx 2.0-3.0$. Notice that the region where the Bloch oscillations take place and the usual bound states are very close to each other, and the amplitude of the Bloch oscillations is weaker compared to the bound states. 
	We zoom in on the Bloch frequencies in \cref{fig:fv}(d) where the overlap between the bound state spectrum and Bloch frequencies is more apparent. And finally, in \cref{fig:fv} (b) we show the line plot for $\chi(t_p,\omega)$ in (c), which again shows the existence of extra frequencies in this setup compared to the true vacuum case.

	To isolate these emergent frequencies, we use a frequency-selective probe to put the bound state spectrum out of focus and zoom into the frequency range of the Bloch frequencies. The frequency selective probe has the following  profile,
	\begin{align}{\label{eq:selective}}
		\phi(t) = A e^\frac{-(t-t_p)^2}{2\sigma^2 }\sin(\omega (t-t_p)),
	\end{align}
	which has frequency support centered around $\omega$ with a width $\sigma^{-1}$.
	By centering our probe around the Bloch frequency with a small width $\sigma$, we can investigate the amplitude and trend of the Bloch oscillations as $|h_l|$ relaxes back to zero.
	
	In \cref{fig:bloch} we apply the functional derivative analysis with the frequency selective probe and focus on the region $\omega\approx1.0-3.5$ where the Bloch oscillations are expected to be. The time and frequency domain profiles of our probe field can be seen in \cref{fig:bloch} (c) and (d).
	In panel (a), we plot $|\chi(t_p,\omega)|^2$ where we focus on the regularly spaced frequencies that lie just below the bound state spectrum. Comparing this to \cref{fig:fv} (c)-(d), we see that these frequencies are clearer in the frequency selective case revealing the dynamics under the time-dependent field, i.e., they disappear as the system re-equilibrates and reaches a low-confinement regime. \Cref{fig:bloch} (b) shows the line plot where the frequency-selective aspect is the most apparent. We see only one peak in most of these lines compared to the numerous peaks in \cref{fig:fv} (b). 
	
	We also show the time-domain dynamics in \cref{fig:bloch} (e) and (f) of $|\delta Z_i(t)|$ for two probe times $t_p\approx17.0$ where we the Bloch oscillations are more prominent and at $t_p \approx 43.0$ where the systems is close to re-equilibration and the Bloch oscillations are not as prominent. We see that the time domain data also agrees with the above expectations. In \cref{fig:bloch} (e), we see that the Bloch oscillations are visible in the time domain data, and the system is confined. However, in \cref{fig:bloch} (f) we see the disappearance of the Bloch oscillations and the full lightcone reappearing in agreement with the frequency domain analysis. 
	
	Our results show that a pump-probe experimental setup could prove useful in tracking the dynamics for a general false-vacuum decay by allowing easy access to the persistent oscillatory behavior that inhibits the excitations from spreading using various probes with various frequency supports. In appendix ~\ref{app:time-domain-gf}, we further investigate the validity of obtaining $\chi^R(t,t_p)$ by doing a Fourier transform of $\chi^R(\omega, t_p)$ back to the time-domain and pin-point the features we see in comparison between $\delta Z(t)$ and the correlation function $\chi^R$.

	Furthermore, while quench spectroscopy is a well-known procedure to study Bloch oscillations, here we show that by simulating a frequency-selective pump-probe experiment, we can extract certain aspects more clearly compared to the usual quench spectroscopy techniques. In particular, one can study the effects of turning on a constant force on electron transport in a lattice; by looking at the frequency range corresponding to the localized wave function and controlling the amplitudes of such Bloch oscillations, transport can be turned on or off subject to different driving fields\cite{hansen_magnetic_2022,song_coherent_2024,verdel_real-time_2020,guo_observation_2021}.

	\section{Conclusions}
	
	In conclusion, we were able to study the low-energy spectrum of the mixed-field Ising model using the functional derivatives approach in the linear response regime. This allowed us to see the confining dynamics arising from a true vacuum initial state, as well as the anti-confining dynamics that play a big part in answering the question of how the false-vacuum state decays in many scenarios. 
	
	Moreover, we have shown that our method is capable of tracking relaxation dynamics of out-of-equilibrium models that do not necessarily exhibit time-translation invariance by setting up simulations or circuits that resemble pump-probe experiments. This is mainly done by introducing a number of probes that take snapshots of the dynamics as time evolves. We also demonstrated the utility of frequency selectivity to zoom in or study emergent peaks in frequency ranges that might not always be easy to access. This allowed us to investigate the emergence of Bloch oscillations and to study their trend in a non-equilibrium setting.  
	
	The power of this result opens up the realm of studying non-integrable models on near-term quantum computers with circuits that do not require ancillary qubits and are robust to noise\cite{kokcu_linear_2024}. With an ancilla-free circuit providing response functions to the first order, one can afterwards post-process over the two-kink states (states with two domain walls) to find the bound-state spectrum in the two-kink model approximation\cite{rutkevich_weak_2010,rutkevich_kink_2018} and many more interesting observables. 
	
	In addition to frequency-selectivity, we can achieve tailored excitations within a certain energy range or of specific momenta using a \textit{momentum-selective} probe. We then obtain response functions directly in the momentum space, making it easier to obtain the electronic structure as is obtained by ARPES experiments.

	Our methods can be further used to study the collision between individual bubbles undergoing inelastic scattering, and the possibility of forming another mesonic state.~\cite{Jordan:2012xnu,PhysRevD.99.094503,Jha:2024jan,Farrell:2025nkx,Bennewitz:2024ixi}. 
	Finally, studying low-energy confining dynamics can open doorways to understand false vacuum decay \cite{lagnese2021false} and string breaking \cite{lerose_quasilocalized_2020} in many applications. 
	As such, we think our work offers a new perspective on simulating non-equilibrium systems and obtaining response functions on both classical and quantum computers.

	\begin{acknowledgments}
		HAL and LFK conceived the initial idea for the work. HAL carried out all the numerical work. GCT contributed to the development of the theoretical background. All authors contributed to the editing and writing of the manuscript. HAL, LFK, and JKF acknowledge financial support from the U.S. Department of Energy, Office of Science, Basic Energy Sciences, Division of Materials Sciences and Engineering under Grant No. DE-SC0023231. JFK received support from the McDevitt bequest at the Georgetown University. GCT acknowledges financial support from the National Science Foundation under award No. 2325080:PIF: Software-Tailored Architecture for Quantum Co-Design. 
	\end{acknowledgments}
	
	\bibliography{confinement}
	
	\clearpage
	\onecolumngrid
	\appendix
	
	\renewcommand\thefigure{A\arabic{figure}}  
	\renewcommand\thetable{A\arabic{table}}  
	\setcounter{figure}{0}
	
	\appendix
	\section{\texorpdfstring{Numerical Extraction of the Response Function \textbf{$\chi^R(t,t_p)$ }}{Analysis}}\label{app:time-domain-gf}
	The analysis in the main text showed the retarded correlation function in the frequency domain $\chi^R(t_p,\omega)$ and $\delta Z_i(t)$ in time-domain. The $t_p$ axis comes in from the pump-probe experimental setup where a stream of probes perturbs the system at various times $t_p$ in separate experiments. As shown in the main text, this allowed us to gain insight into the relaxation dynamics as the pump effect dissipates and the system reaches equilibrium. In this section, we further investigate the effect of the different shapes of the probe pulses on the extracted response function $\chi^R(t,t_p)$. While the observable expectation value is related to the response function $\chi^R(t,t_p)$ in the following way,
	\begin{align}
		\delta Z(t) = \int \chi^R(t,t') \phi(t') dt'
	\end{align}
	the convolution between the response function and probe pulse adjusts the dynamics in both time and frequency domains.
	We address the conservation of causality and the validity of obtaining $\chi^R(t,t_p)$ from $\chi^R(t_p,\omega)$ by an inverse Fourier transform in the broadband probe pulse and a frequency-selective probe pulse both under the false vacuum simulation tackled in the main text.

	\begin{figure}[H]
		\centering
		\includegraphics[width=1\linewidth]{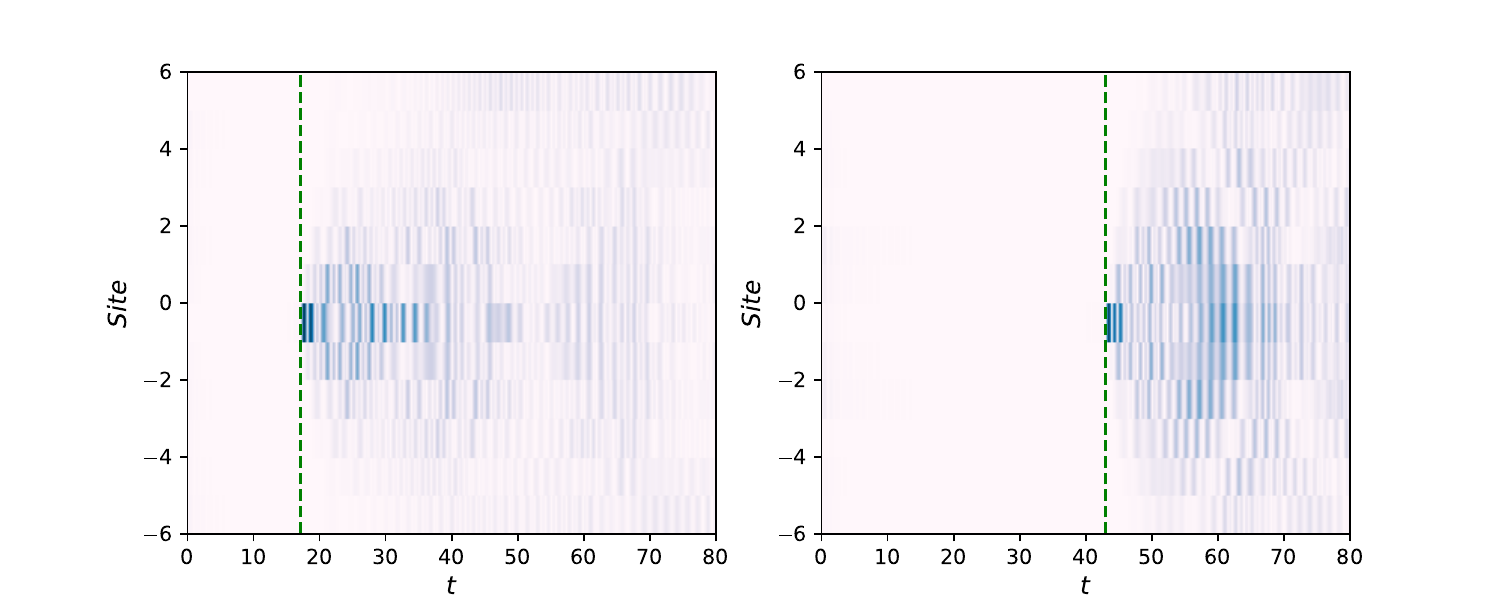}
		\caption{ \textbf{$\chi^R(t,t_p)$ with a Gaussian probe pulse. }$\chi^R(t,t_p)$ obtained at two different times $t_p=(17.12, 42.92)$ along the $t_p$ axis indicated by the green dashed lines in both panels. The probe is a narrow Gaussian of $\sigma = 0.2$. We show here that causality is conserved and $\chi^R(t,t_p)$ is easily obtained by performing a Fourier transform back to time domain.}
		\label{fig:appendix_gauss}
	\end{figure}
	\subsection{False Vacuum Analysis: A Broadband Probe}
	In \cref{fig:fv}, the analysis probed the full low-energy spectrum using a Gaussian probe that supports a wide range of frequencies. While $\delta Z_i(t)$ and $\chi^R(t,t_p)$ carry similar information, $\chi^R(t,t_p)$ is the correlation function to linear order that tells more about the response of the system to the perturbation and from which we can extract the spectrum. \cref{fig:appendix_gauss} shows $\chi^R(t,t_p)$ at two different probe times. The first panel shows an early time-point where the dynamics are in the high-confinement regime; while the second panel shows a later time-point when the system reached low-confinement. We see this from the propagation of information over the site axis in both cases. Causality is fundamentally embedded in the structure of the response function $\chi^R(t,t_p) $ which satisfies 
	\begin{align}
		\chi^R(t,t_p) = \theta(t-t_p)\langle \psi_0| [Z(t),Z(t_p)]|\psi_0\rangle.
	\end{align}
	This ensures that the system's response at time
	$t$ cannot depend on a perturbation applied at a later time $t_p$. Since the probe is chosen to be a narrow Gaussian, we expect no signal before $t=t_p$ that may arise from a non-compact nature of the probe.

	\begin{figure}[h]
		\centering
		\includegraphics[width=1.0\linewidth]{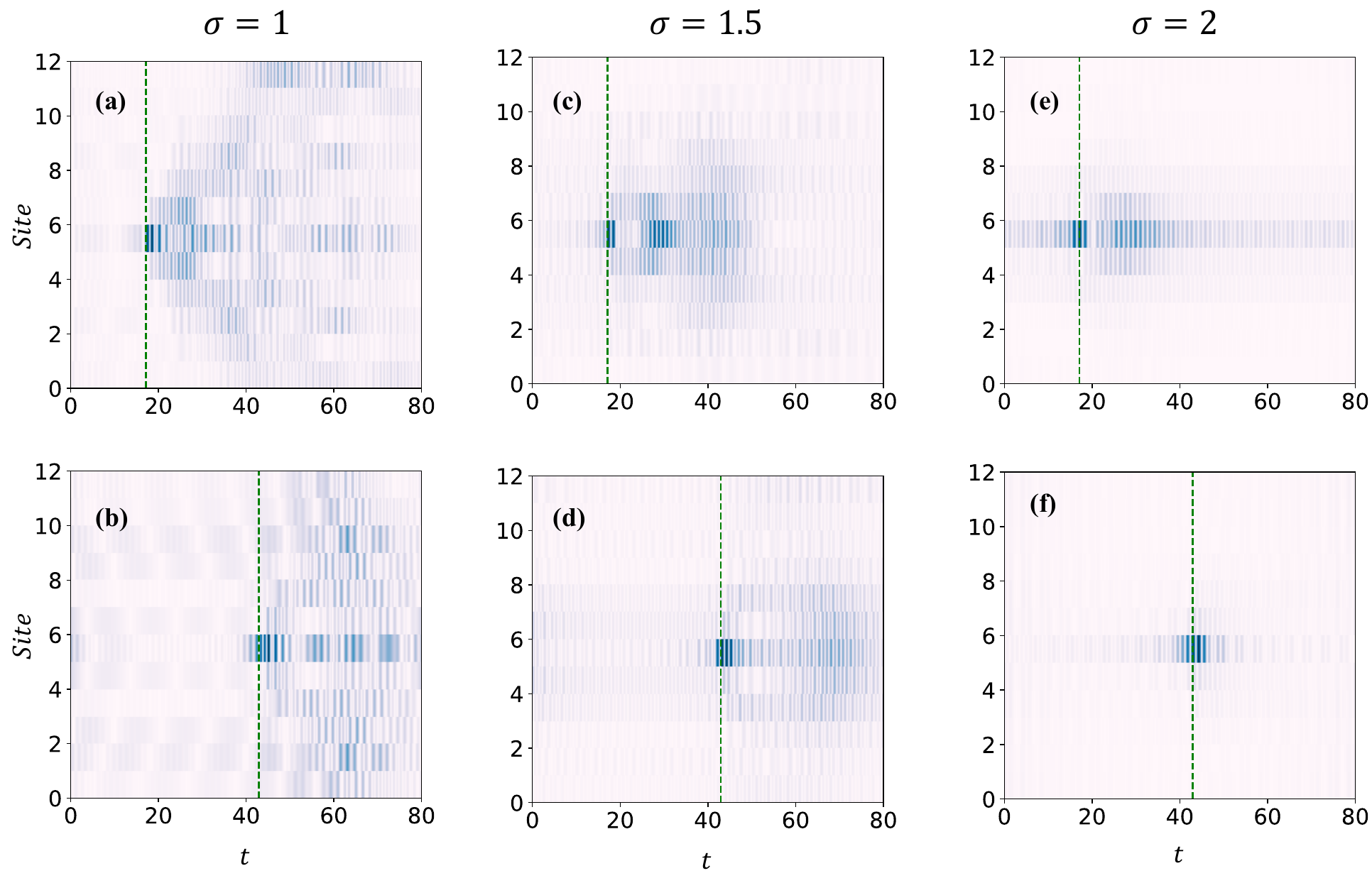}
		\caption{\textbf{$\chi^R(t,t_p)$ with a Frequency-Selective probe pulse. }$\chi^R(t,t_p)$ obtained at two different times $t_p=(17.12, 42.92)$ indicated by the green dashed lines in all panels and at 3 different frequency-selective probes centered at $\omega = 2.2$ and of various widths $\sigma$. Panels \textbf{(a)} and \textbf{(b)} show the correlation function with a frequency-selective probe of $\sigma =1$,  \textbf{(c)} and \textbf{(d)} show the correlation function with a frequency-selective probe of $\sigma =1.5$, and \textbf{(e)} and \textbf{(f)} show the correlation function with a frequency-selective probe of $\sigma =2$. The figure aims to pinpoint the different background structures before $t_p$ for different probe profiles and emphasize the idea of causality while enforcing the fact that a frequency-selective probe, while it can give support before $t=t_p$, it still preserves it. }
		\label{fig:selective_analysis}
	\end{figure}

	\subsection{False Vacuum Analysis: A Frequency-Selective Probe}
	
	The picture slightly changes when $\chi^R(t,t_p)$ is subject to a frequency-selective probe; a frequency-selective probe, in its essence, is broad in time to ensure a selective nature in the frequency-domain. As such, the selection of a narrow frequency band picks up certain components of the response function to amplify and reduce/omit the rest. In addition to this, the probe of the profile shown in \cref{eq:selective} gives rise to oscillations over the $t_p$ axis $e^{i\omega t_p}$. These altogether may alter the shape of $\chi^R(t,t_p)$ to show oscillations before $t_p$. 
	
	In efforts to show the effect of a frequency-selective probe, even though the dynamics still conserves causality, we compare probes of different widths $\sigma$ and obtain the retarded correlation function after an inverse Fourier transform in \cref{fig:selective_analysis}. The objective of \cref{fig:selective_analysis} is to show the different noise/background structures subject to probes of different widths, as well as showing how the signal still slightly starts before $t_p$ in proportion to the width $\sigma$. 
	
	We de-modulate the signal by applying a $e^{i\omega t_p}$ before inverse Fourier transforming. This is successful in removing the overall envelope $\sin(\omega t_p)$, but does not change how the remaining components of the signal sum up to give a $\chi^R(t<t_p) \neq 0$. We emphasize that this does not break causality, and the quantity measured is still a retarded correlation function.
	
	In panels (a) and (b) of Fig~\ref{fig:selective_analysis}, the width of the probe is set to be $\sigma = 1$. The background structure before $t=t_p$ reflects the components of the signal that remain within the bandwidth of the probe. In panels (c) and (d), we show the exact thing, but with a slightly larger probe width of $\sigma=1.5$, this covers a smaller frequency range, and thus a different background structure is built. Similarly, for panels (e) and (f), we find a similar situation with an even smaller bandwidth in the frequency domain. We point up that the signal after $t_p$ would only show the oscillations in the response function pertaining to one of the frequencies allowed by the selective probe. This is one of the powers of utilizing such a pulse in experimental and computational settings.

	\section{Dynamics Along the $t_p$ Axis}\label{app:tp}
	
\begin{figure}
		\centering
		\includegraphics[width=1.0\linewidth]{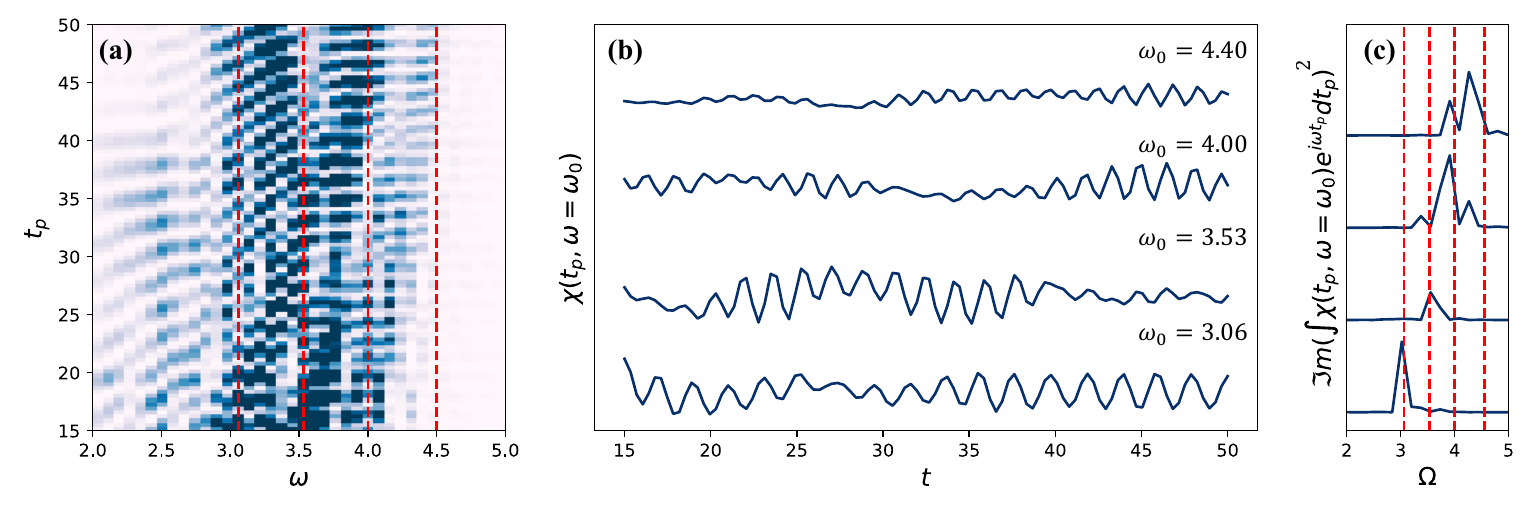}
		\caption{\textbf{Dynamics along the probe axis $t_p$. }\textbf{(a)} $\chi(t_p, \omega)$ in the false vacuum case. Red lines indicate the vertical cuts we take at specific $\omega=\omega_0$ frequencies. \textbf{(b)} The dynamics of $\chi(t_p,\omega=\omega_0$ at four different values as a function of $t_p$. \textbf{(c)} The Fourier transform along $t_p$ axis for each $\chi(t_p,\omega=\omega_0)$. The figure aims to further investigate the oscillations along the $t_p$ axis and show the reflection of the mesonic masses on the probe axis $t_p$. We find that the specific mesonic mass where the vertical cut is made is also reflected along it. }
		\label{fig:tp-dynamics}
	\end{figure}

	In this section, we aim to explore the dynamical structure along the probe time axis $t_p$. By applying a weak pulse at various times $t_p$, and under a time-dependent Hamiltonian due to the pump at $t=0$, the response $\chi^R(t,t_p)$ depends non-trivially on both $t$ and $t_p$. In such a scenario, the excitation spectra can be reflected along the different axes; if we probe the dynamics along $t_p$, we are probing how long the system retains memory of the perturbation and how it relaxes back to equilibrium. In \cref{fig:tp-dynamics}, we show the $\chi(t_p)$ obtained by Fourier transforming $\chi(t_p,\omega=\omega_0)$ along $t_p$. As such, we start by taking a vertical cross-section at $\omega = \omega_0$ through the false color map we show in \cref{fig:tp-dynamics}(a) such that we achieve
	\begin{align}
		\chi^R(t_p,\omega=\omega_0) 
	\end{align}
	This analysis helps us to study the evolution of specific frequencies along the $t_p$ axis. In panel (b), we look into the time dynamics of the vertical cross-sections along the same time axis $t_p$ and then, in panel (c), we perform a Fourier transform 
	\begin{align}
		\Im m(\int \chi^R(t_p,\omega=\omega_0) e^{i\omega t_p} dt)^2 
	\end{align}
	Doing this, we find the reflection of each specific mesonic mass on the vertical axis.

\end{document}